\magnification=1200
\baselineskip=18pt
\hsize=12 cm
\vsize=18 cm
\font\bbold=cmbx10 scaled\magstep2
\centerline{\bbold The Spin-S Blume-Capel RG Flow Diagram} 
\vskip 5pt
\centerline{S. Moss de Oliveira and P.M.C. de Oliveira}\par
\vskip 3pt
\centerline{Institute of Physics - Fluminense Federal University}\par
\centerline{Cx. Postal 100296, Niter\'oi, RJ, Brazil 24001-970}\par
\centerline{e-mail GFIPMCO@BRUFF.bitnet or GFIPMCO@VMHPO.UFF.BR}
\vskip 5pt
\centerline{F.C. de S\'a Barreto}
\vskip 3pt
\centerline{Departamento de F\'{\i}sica ICEX - Universidade Federal de Minas Gerais}\par
\centerline{Cx. Postal 702, Belo Horizonte, MG, Brazil 30161-970}\par
\centerline{e-mail SBARRETO@BRUFMG.bitnet or SBARRETO@FISICA.UFMG.BR}\par
\vskip 0.5cm \leftskip=1 cm \rightskip=1 cm
\noindent {\bf Abstract}\par
Using the Finite Size Scaling Renormalisation Group we obtain the
two-dimensional flow \ \  diagram\ \   of\ \   the\ \   Blume-Capel\
\   model, for $S = 1$
and $S = 3/2$. In the first case our results are similar to those of
Mean Field Theory, which predicts the existence of first and second
order transitions whith a tricritical point. In the second case,
however, our results are different. While we obtain, in the $S = 1$
case, a phase diagram presenting a multicritical point, the Mean
Field approach predicts only a second order transition and a critical
end point.\par   
\vskip 0.5cm
\noindent Work partially supported by Brazilian agencies FINEP, CNPq, CAPES and
FAPEMIG.
\leftskip=0 pt \rightskip=0 pt
\vskip 10pt  
\noindent {\bf I - Introduction}\par
The Finite Size Scaling Renormalisation Group (FSSRG) was introduced in 
[1], and it is based only on the finite
size scaling hypothesis [2], with no further assumptions. Its main idea is to
construct particular quantities scaling as $L^0$ in the thermodynamic limit
$L \to \infty$, $L$ being the linear size of the system. By
preserving these quantities, independently of particular 
prescriptions relating states $S$ of the largest lattice to $S'$ of the
smallest one, one obtains RG recursion relations without any conceptual
problem. The construction of these quantities is based on
the symmetries between the various ground states of the system.
The FSSRG in its first version [1] was applied to study the Ising model in
hypercubic lattices. More recently, using it to study the three-dimensional
diluted Ising model, we were able to find a semi-unstable fixed point in
the critical frontier concentration $p$ versus exchange coupling $J$ [3],
characterizing a universality class crossover when one goes from pure to
diluted Ising ferromagnets. Moreover, the specific heat exponents we obtained
for the pure and diluted regimes are in agreement with the Harris
criterion [4]. Based on such a success, we decided to use the FSSRG to study
the two dimensional spin-$S$ Blume-Capel Model.\par
The model is a simple generalization of the spin-1/2 Ising model, described by
the Hamiltonian
$${\cal H} = -\,\,J \sum_{<\imath\jmath>} S_\imath S_\jmath + D \sum_\imath
{S_\imath}^2 \ \ \ \ ,
\eqno (1)$$
where $J > 0$ and the spins $S_\imath$ have values
$-S,\,\,-S+1,\,\,...\,\,,S$, with $2S$ being an integer. The first
sum is over all nearest-neighbor sites of the lattice and $D$ is the
parameter of anisotropy 
(single-ion anisotropy). For $S = 1$, ${\cal H}$ is a well studied model
introduced by Blume [5] and Capel [6] to describe critical-tricritical
phenomena. On the temperature-anisotropy phase diagram the ordered
ferromagnetic and the disordered paramagnetic regions are separated by a phase
boundary which changes character at a tricritical point. According to mean
field calculations the first order transition line goes to zero at $D/zJ =
1/2$, where z is the coordination number. This result can also be
exactly obtained by simple reasonings about the ground states (see
section {\bf II}). Several approximate schemes have been
used to stablish the thermodynamic behavior of this model (see references in
[7]). For arbitrary spin $S$, the model has been less studied. A recent mean
field solution [7] of the general spin-$S$ Blume-Capel model shows that for
integer spins there exist one tricritical point and a disordered phase at low
temperature which are not present for semi-integer spins.
There exists at $T = 0$, a multiphase point from
which a number of first order boundaries spread out. In particular, for the
spin-3/2 model there is one first order line ending at an isolated critical
point, inside the ordered ferromagnetic region. In the following section we
analyse the model for $S = 1$. The various phase regions are
qualitatively determined. In section III, we introduce the method. A more
detailed description can be found in reference [3]. In section IV we present
the results for the flow diagram for the $S = 1$ and $S = 3/2$ models in the
two dimensional square lattice. Finally, in section V, we present some
concluding remarks. 
\vskip 10pt  
\noindent {\bf II - Analysis for $S = 1$}\par
Before explaining the FSSRG method, it is interesting to investigate which
informations can be extracted in advance from the Hamiltonian. We start
rewriting eq.(1) as
$$-{\cal H} = J \sum_{<\imath\jmath>} S_\imath S_\jmath + {\cal D}
\sum_\imath \sigma_\imath\ \ \ \ , 
\eqno (2)$$
where $S_\imath$ = +1, 0 or $-1$, $\sigma_\imath = 2{S_\imath}^2 - 1$ = +1
or $-1$ and ${\cal D} = -D/2$. It is easy to verify that these two
Hamiltonians are equivalent up to an additive constant. The following
informations can be obtained from eq.(2):
\vskip 5pt
\noindent {\bf i)} The symmetry $J \leftrightarrow -J$\par
\noindent Dividing the lattice in two sublattices (chessboard), inverting all
the spins of one sublattice and changing the sign of $J$, the system remains
invariant. That is, the ferromagnetic order is symmetric to the
antiferromagnetic one. Therefore, it is enough to study the $J \geq 0$ half of
the phase diagram.\par
\vskip 5pt
\noindent {\bf ii)} The limit when the temperature $T \rightarrow 0$
with ${\cal D} < 0$\par  
\noindent In this limit the system is in one of its three possible ground
states, one corresponding to all spins $S_\imath = 1$, the other to all
$S_\imath = -1$ and the third one to all spins $S_\imath = 0$. For the first
and second ground states the Hamiltonian per site can be written as
$$-{\cal H}_F/N = 2J + {\cal D} \ \ \ \ , 
\eqno (3)$$
where the factor 2 corresponds to the ratio (number of bonds)/(number of
sites) for a square lattice. For the third ground state (all $S_\imath = 0$)
the Hamiltonian per site is given by:
$$-{\cal H}_Z/N = - {\cal D} \ \ \ \ . 
\eqno (4)$$
It is important to understand that the two former ground states correspond to
an usual ordered ferromagnetic phase with a nonzero magnetisation. The third
ground state, however, corresponds to an ordered phase but with zero
magnetisation, with the majority of the spins equal to zero. A first order
transition will occur for ${\cal H}_F = {\cal H}_Z$, that is
$$2J + {\cal D} = -{\cal D} \ \ \ \ \Rightarrow \ \ \ \ J + {\cal D} = 0 \ \ \ \ . 
\eqno (5)$$
In this way the phase diagram $J/T$ versus ${\cal D}/T$, in the limit $J, -{\cal D} \rightarrow
 \infty$ corresponds to a straight line given by $J/T +
{\cal D}/T = 0$. For $(J/T + {\cal D}/T) > 0$ the system is in the usual Ferromagnetic
phase, and for $(J/T + {\cal D}/T) < 0$ it is in the ``Zero phase''
mentioned above. This result is also reproduced within mean field
theory [5,6,7]. 
\vskip 5pt
\noindent {\bf iii)} The limit ${\cal D} \rightarrow \infty$ \par
\noindent In this\ \   case\ \   $S_\imath$ = +1 or $-1$,\ \   and
on\ \   this Ising limit we
shall\ \  find $J_c = (1/2) \,\, \ln (\root \of 2 +1) = 0.4407$ and $\nu =
1$ for the critical coupling and correlation lenght exponent,
respectively. Hereafter, we take $T = 1$ for simplicity.
\vskip 5pt
\noindent {\bf iv)} The limit $J \rightarrow 0$ \par
\noindent In this limit the spins are independent and we can write
$$Z = {Z_\imath}^N = {(2 \,\, e^{\cal D} + e^{-{\cal D}})}^N \propto
{(e^{{\cal D} \,\, +
\,\, (1/2) \,\, \ln 2} + e^{-{\cal D} \,\, - \,\, (1/2) \, \, \ln 2})}^N
\ \ \ \ .$$  
Therefore there is a ``transition'' from the Zero phase to a
Paramagnetic phase when ${\cal D} + 1/2 \,\, \ln 2 = -{\cal D} - 1/2
\,\, ln2$, that is ${\cal D}_c = -1/2 \,\, \ln 2 = 
-0.3466$. This is a phase transition induced by an external
field, and corresponds to no singularities in the thermodynamic quantities.
Since both phases have zero magnetisation, they cannot be
distinguished through a mean field approach which usually measures the
magnetisation to identiffy the different phases of the system.
\vskip 10pt
\noindent {\bf III - The Method}\par
One of the most important advantages of the FSSRG is that one does not need 
to adopt any particular recipe of the type
$$\exp(-{\cal H}'(S')/T) = \sum_S P(S,S') \exp(-{\cal H}(S)/T) \ \ \ \ ,$$ relating the
spin states $S$ of the original system to the spin states $S'$ of a
renormalised system. In traditional Real Space Renormalisation Group (RSRG)
implementations, the choice of a particular weight function $P(S,S')$, e.g. the
so called majority rule, is generally based on plausibility arguments, and
involves uncontrollable approximations. Based on this weakness, many
criticisms have been made about RSRG since its introduction twenty years ago
[H]. In spite of these criticisms, RSRG has been an important investigation
tool since then. 
On the other hand, FSSRG is free from these criticisms and shares with RSRG
some good features as, for instance, the possibility of extracting {\it
qualitative} informations from {\it multiparameter RG flux diagrams}, including
crossovers, universality classes, universality breakings, multicriticalities,
orders of transitions, etc. Other unpleasant consequences of particular weight
functions, as the so called proliferation of parameters, are absent from
FSSRG.\par 
In order to explain the method we will consider a $d$ dimensional
hypercubic lattice with its cover and bottom $d-1$ dimensional
hypersurfaces. There are two quantities to be preserved  
in the renormalisation process: the magnetic quantity ${\cal Q}$ and the
thermal quantity ${\cal T}$. The first one is defined by
$${\cal Q} = \biggl \langle {\rm sign} \bigl ( \sum_{\imath} \sigma_\imath
\bigr ) \biggr \rangle  \ \ \ \ ,
\eqno (6)$$ 
where $\langle ... \rangle$ means the
canonical thermodynamic average, and sign$(x)$ = $-1$, 0 or $+1$ for $x < 0$,
$x = 0$ and $x > 0$,  respectively.
This quantity is referred as a magnetic quantity because it has a nonzero 
value only in the presence of a magnetic field (which in this model is the
field ${\cal D}$). In this way it is related to the symmetry breaking of the
system. The other quantity is defined, through the surfaces of the lattice, as
$${\cal T} = \biggl \langle \mu (top) \,\, \mu (bottom) \biggr
\rangle \ \ \ \ ,  
\eqno (7)$$
where $\mu = +1$ if the majority of the spins $S_\imath = +1$, $\mu = 0$ if
the majority of the spins $S_\imath = 0$ and $\mu = -1$ if the majority of the
spins $S_\imath = -1$. That is, the quantity ${\cal T}$ is related to the
sign of the majority of the spins of the top and bottom hypersurfaces.
There are three possible values of $\mu$ because there are also three
different possible ground states. If we were studying the Ising Model
($S_\imath = \pm 1/2$), for instance, we should admit only two possible values
for $\mu$. As extensively explained in ref. [3], the quantity ${\cal T}$ is
related to ergodicity or long range order breaking. It vanishes above $T_c$
(paramagnetic phase) and equals 1 below $T_c$ (ordered phase)
independently of the existence of a magnetic field. Another point which is
also carefully discussed in ref. [3] is the zero value of the anomalous
dimension $\phi$ of both ${\cal Q}$ and ${\cal T}$, which is a
consequence of their definition. Just because $\phi = 0$ we can guarantee that
these quantities scale as $L^0$ and so are preserved during the renormalisation
process. 
\vskip 10pt
\noindent {\bf IV - The Lattices and Results}\par
\vskip 5pt
\noindent {\bf Results for $S = 1$}\par
In Fig.1 are shown the lattices we have used. In fact, these are the
smallest ones that could be chosen for a renormalisation process. Usually the
best way to respect the ratio (number of bonds)/(number of sites) = 2 of the
infinite square lattice, when using finite lattices approximations, is to adopt
periodic boundary conditions. However, for such tiny lattices, it is more
convenient to attribute weights to the sites according to their coordination
number [9] as shown in Fig.1. The corresponding renormalisation group equations
are of the form 
$${\cal Q}_{L'} (J',{\cal D}') = {\cal Q}_L (J,{\cal D})
\eqno (8a)$$
and
$${\cal T}_{L'} (J',{\cal D}') = {\cal T}_L (J,{\cal D}) \ \ \ \ ,
\eqno (8b)$$
where the primes are referred to the smallest lattice of size $L'$.\par
In Fig.2 it is shown the resulting flow diagram. The attractors of the
different phases are represented by squares, critical points by dots
and the tricritical point by a star. The full line is a second
order critical frontier, the dashed line is a first order one and the
dotted line, as mentioned before, 
separates the usual paramagnetic phase from the ``zero phase''.\par
The critical point at the upper part of the diagram (${\cal
D} \rightarrow \infty$) is related to the analysis made in section {\bf
II-iii)}, and at this point we have found $J_c = 0.4730$. It is
important to note that we are mainly interested in the qualitative features
of the phase diagrams, and that with so tiny lattices it would not
make sense to compare our numerical results to those obtained for
instance, through Monte Carlo calculations or conformal invariance
arguments [10]. Nevertheless, any predefined
numerical accuracy can be 
obtained through the present method, simply by using large enough
lattices. One can, for instance, calculate the quantities ${\cal Q}$
and ${\cal T}$ through Monte Carlo sampling [1].

The critical point at $J = 0$ is related to section {\bf II-iv)},
and we have found ${\cal D}_c = 0.3478$. At the tricritical point
we have obtained $J_T = 3.5836$ and ${\cal D}_T = -3.5830$, in
agreement with the analysis made in section {\bf II-ii)}.\par
Considering $b$ the scaling factor and $d$ the dimension of the
system, a first order transition can be characterized by the magnetic
eigenvalue $\lambda = b^d$
calculated at the attractor of the critical line, that is, at the
critical point that can be seen at the lower right corner of our
phase diagram ($J, \,\, -{\cal D} \rightarrow \infty$). Taking the
limit $T \rightarrow 0$ $({\cal D}< \,\,0)$ 
in our RG equations we have found $\lambda = 4 = b^d$, which ensures
that the dashed line represents in fact a first order transition. 
\vskip 10pt
\noindent {\bf Results for $S = 3/2$}\par
In this case the Hamiltonian can also be represented by eq.(2), but
now with $\sigma_\imath = ({S_\imath}^2 - 5/4)$ = +1 or $-1$ for $S_\imath =
\pm 3/2$ and $S_\imath = \pm 1/2$, respectively. The same analysis
made for the $S_\imath = 1$ case can be easily performed again, and
also the same calculation method can be applied. The resulting RG
flow diagram is shown in Fig.3. The phases $F_1$ and $P_1$ are
related to $S_\imath = \pm 3/2$, and $F_2$ and $P_2$ to $S_\imath =
\pm 1/2$. Again for the magnetic eigenvalue calculated at the lower right
critical point of the diagram we have $\lambda = 4 = b^d$, indicating
the first order character of the dashed critical line. At the upper
left critical point (${\cal D} \rightarrow \infty$) we have found
$J_c = 0.0525$, and 
at the $J = 0$ critical point, ${\cal D}_c = 0$. At the
tetracritical point we have $J_T = 1.0967$ and ${\cal D}_T = - 8.7831$. The
lower critical point obtained for ${\cal D} \rightarrow - \infty$ is
again the Ising spin-1/2 Onsager value, and we get $J_c = 0.4730$.
\vskip 10pt
\noindent {\bf V - Concluding Remarks} \par
In summary, we have calculated the flow diagram of the general spin-$S$
Blume-Capel Model, for $S = 1$ and $S = 3/2$, by means of the FSSRG.
For the $S = 1$ case the results reproduce previous calculations
based on mean-field approximation as well as other approaches (see
[7] and references therein). For the spin-3/2 case new features are
presented. The flow diagram for this case, as far as we know, has
never been shown before. Besides, contrary to mean field 
results [7], we show the appearence of a tetracritical point in the phase
diagram, instead of an isolated critical end point. We remark that
the numerical values obtained in these calculations are 
not accurate enough, due to the fact we have used small size lattices to
implement the renormalisation process. However, the results are
expected to be qualitatively
correct and improved quantitative values can be obtained by the use of larger
lattices cells. Finally, the same procedure can be applied for other values of
spin-$S$. We expect similar qualitative behavior as obtained here.\par
\vfill\eject
\noindent {\bf References}\par
\item{[1]} de Oliveira P.M.C., {\it Europhysics Letters} {\bf 20}, 621
(1992).\par 
\item{[2]} Fisher M.E., in {\sl Proceedings of the International Summer School
Enrico Fermi, Course {\rm LI} Critical Phenomena}, ed. Green M.S., Varenna,
Italy (Academic Press, 1971).\par
\item{[3]} de F. Neto J. Monteiro, de Oliveira S. Moss and de Oliveira P.M.C.,
{\it Physica} {\bf A 206}, 463 (1994)\par
\item{[4]} Harris A.B., {\it J. Phys.} {\bf C7}, 1671 (1974).\par 
\item{[5]} Blume M., {\it Phys. Rev.} {\bf 141}, 517 (1966).\par
\item{[6]} Capel H.W., {\it Physica} {\bf 32}, 96 (1966); {\bf 33}, 295 (1967);
{\bf 37}, 423 (1967).\par
\item{[7]} Plascak J.A., Moreira J.G. and S\'a Barreto F.C., {\it Phys. Lett.}
{\bf A173}, 360 (1993).
\item{[8]} Niemeijer Th. and van Leeuwen J.M.J., {\it Phys. Rev. Lett.} {\bf
31}, 1411 (1973).\par
\item{[9]} dos Santos R.R., {\it J. Phys.} {\bf C 18}, L-1067 (1985).\par
\item{[10]} Sen D., {\it Phys. Rev.} {\bf B44}, 2645 (1991); Malvezzi
A.L., {\it Braz. J. Phys.} {\bf 24}, 508 (1994).\par
\vfill\eject
\noindent{\bf Figure Captions}\par
\item{Figure 1--} The lattices used in the renormalisation process and the
weights atributed to the sites. Also the cover and botton surfaces
are indicated for this square lattice case.\par
\item{Figure 2--} The FSSRG flow diagram for $S = 1$.\par
\item{Figure 3--} The FSSRG flow diagram for $S = 3/2$.\par
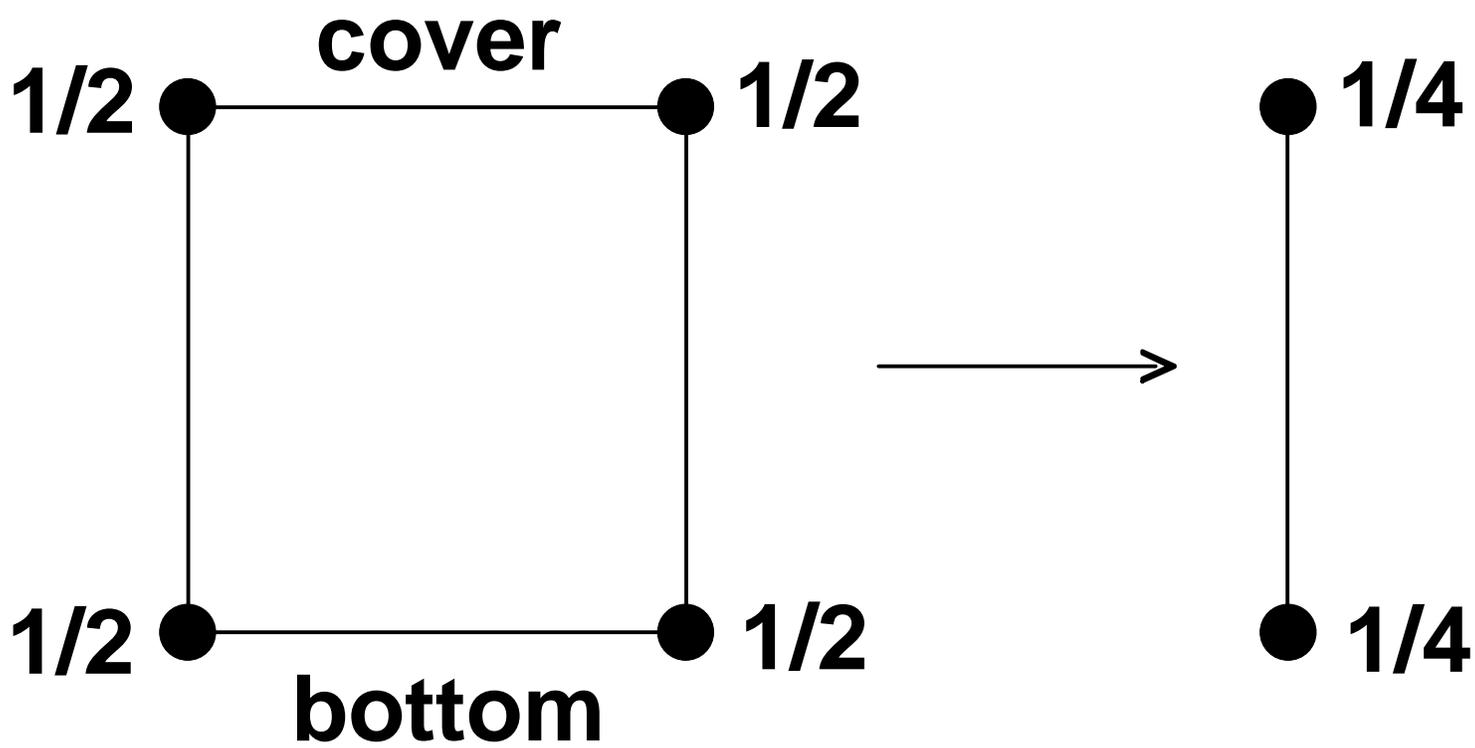
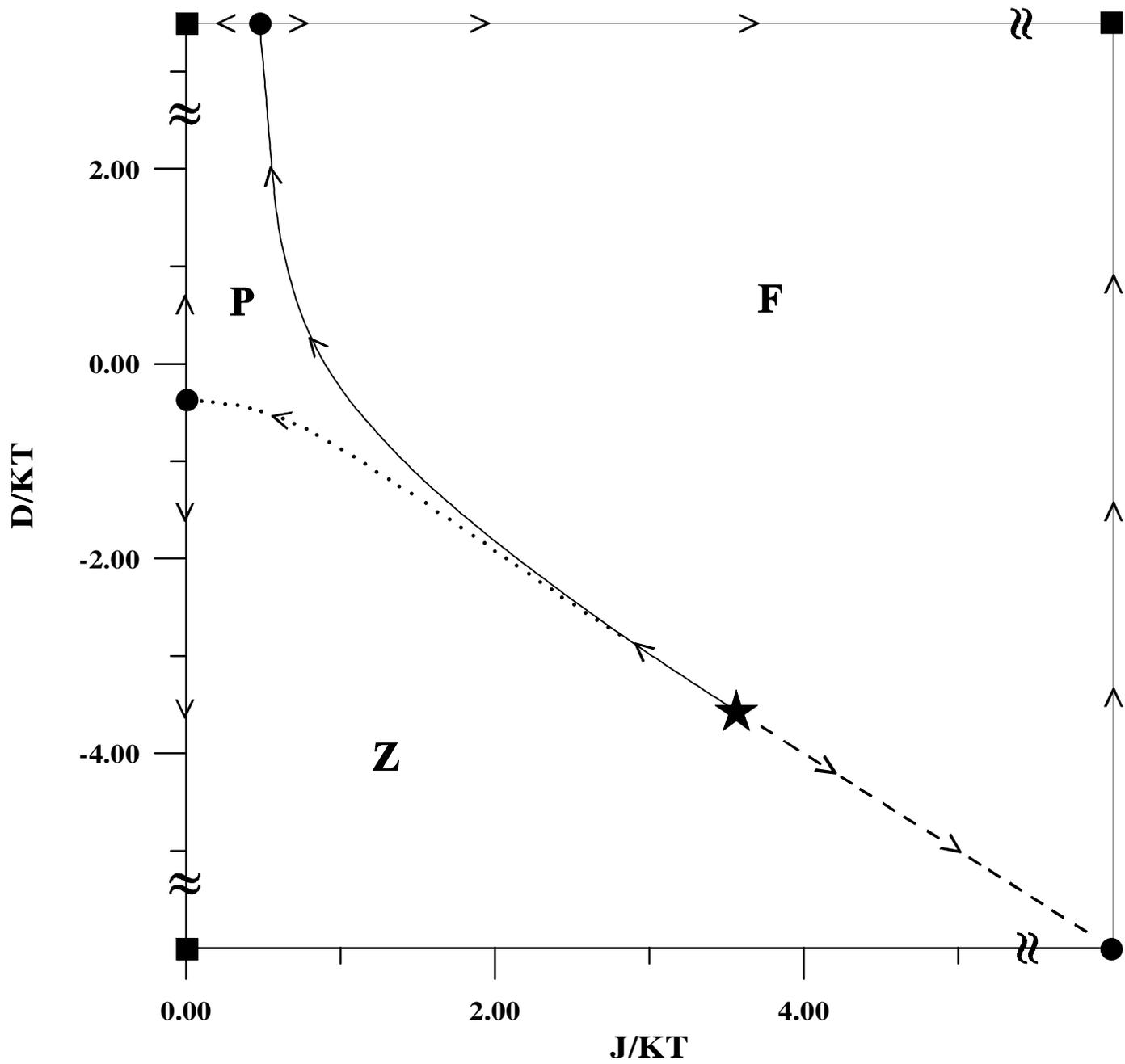
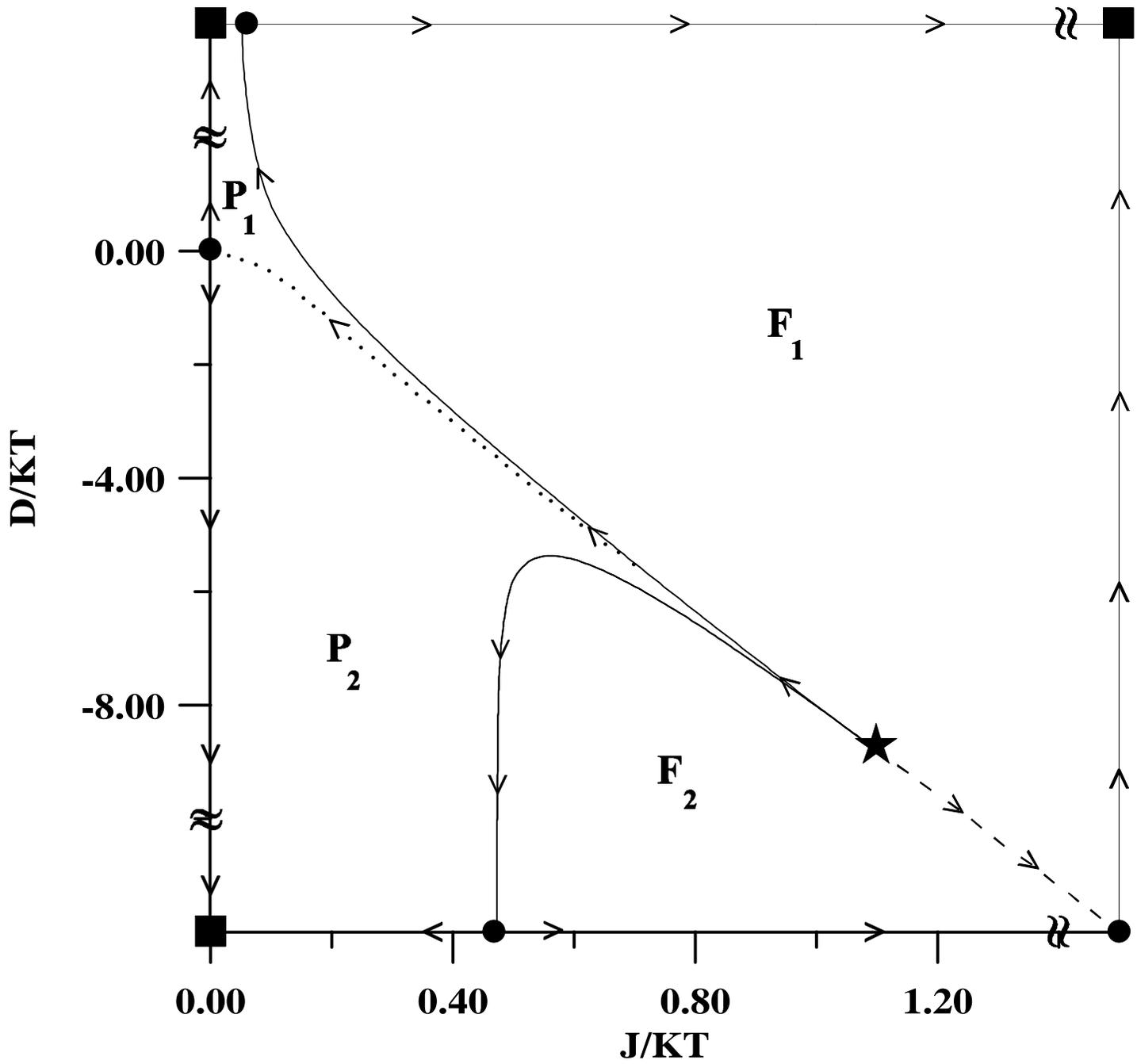
\bye